\documentclass[final,5p,times,twocolumn]{elsarticle}

\usepackage{amssymb}
\usepackage{amsmath}

\usepackage{array,multirow,makecell} 

\begin{document}

\begin{frontmatter}

\title{An experimental setup for the study of gas-cell processes for the S$^3$-Low Energy Branch}

\author[IJCLab]{E. Morin} 
\author[IJCLab]{W. Dong\fnref{Wenling}} 
\author[IJCLab]{V. Manea\corref{ref1}} 
\author[Leuven]{A. Claessens} 
\author[GANIL]{S. Damoy} 
\author[Leuven]{R. Ferrer} 
\author[IJCLab]{S. Franchoo} 
\author[GANIL]{S. Geldhof} 
\author[IJCLab]{T. Hourat} 
\author[Leuven]{Yu.~Kudryavtsev} 
\author[GANIL]{N. Lecesne} 
\author[GANIL]{R. Leroy} 
\author[IJCLab]{D. Lunney} 
\author[IJCLab]{V. Marchand} 
\author[IJCLab]{E. Minaya Ramirez} 
\author[Mainz]{S. Raeder} 
\author[IJCLab]{S. Roset} 
\author[LPC]{Ch.~Vandamme} 
\author[Leuven]{P.~Van den Bergh} 
\author[Leuven]{P. Van Duppen} 

\affiliation[IJCLab]{organization={Université Paris-Saclay, CNRS/IN2P3, IJCLab},
            city={Orsay},
            postcode={91405}, 
            country={France}}
\affiliation[Leuven]{organization={KU Leuven, Instituut voor Kern- en Stralingsfysica},postcode = {B-3001},city ={Leuven}, country = {Belgium}} 
\affiliation[GANIL]{organization={GANIL, CEA/DRF-CNRS/IN2P3},postcode = {14076},city ={Caen}, country = {France}}
          
\affiliation[Mainz]{organization={Institut für Physik, Johannes Gutenberg-Universität Mainz},postcode = {55128 },city ={Mainz}, country = {Germany}}  
\affiliation[LPC]{organization={Université de Caen Normandie, ENSICAEN, CNRS/IN2P3},postcode = {F-14000},city ={Caen},country = {France}}
\fntext[Wenling]{This work contains part of the PhD thesis work of Wenling Dong}
\cortext[ref1]{Corresponding author: vladimir.manea@ijclab.in2p3.fr}

\begin{abstract}
We present an experimental setup dedicated to the study of in-gas ion processes and characterization of gas stopping cells for the  Low Energy Branch of the Super Separator Spectrometer (S$^3$) at SPIRAL2-GANIL. The first application is the development of a new gas stopper with a neutralization mechanism designed for faster extraction of the radioactive ions. This development should enable in-gas-jet laser spectroscopy and other low-energy experiments with shorter lived radioactive isotopes. We discuss in detail the motivation and objectives of these developments and we present the results of simulations performed in the design phase, as well as the first experimental results. 
\end{abstract}


%


\end{frontmatter}



\section{Introduction}
\label{intro}

The SPIRAL2 facility is a new-generation research infrastructure devoted to experiments with radioactive ion beams at GANIL \cite{SPIRAL2}. It is powered by a linear accelerator which can deliver heavy ion beams with intensities above 1 p$\mu A$ and energies between 0.75 and 14.5 MeV/u. One of the key physics programs is the study of neutron-deficient and (super-)heavy nuclei with the Super Separator Spectrometer (S$^3$) \cite{Dechery2015}. To this end, the heavy ion beams induce fusion-evaporation reactions in a rotating target at the entrance of the spectrometer. The reaction products are separated from unreacted primary beam and other contaminants by the spectrometer and delivered to a gas cell located at the S$^3$ focal plane, marking the beginning of the S$^3$ Low Energy Branch (S$^3$-LEB) \cite{Dechery2015, Ferrer2013}. 

The products enter the gas cell through a thin window and are stopped in argon buffer gas at a pressure of 200-500~mbar, where they are thermalized and neutralized. The gas flows out of the cell through a de Laval nozzle, carrying with it the radioactive species \cite{Zadvornaya2018}. The gas expansion through the nozzle leads to the formation of a supersonic gas jet which, owing to the nozzle contour and matched pressure conditions in the expansion chamber, can be highly collimated and homogeneous \cite{Ferrer2021}. In the jet, pulsed lasers are overlapped with the radioactive species to produce step-wise resonance ionization. The photo-ions thus created are extracted by a series of radiofrequency quadrupole (RFQ) ion guides \cite{Major2005}, bunched and delivered to a multi-reflection time-of-flight mass spectrometer referred to as PILGRIM \cite{Chauveau16} or to a decay station called SEASON \cite{SEASON,ReyHerme2023}. 

Scanning the frequency of the first laser step used for excitation and monitoring the ionization rate, either by counting the ions directly or their decay radiation in a given time, allows performing laser spectroscopy measurements.  The adiabatic expansion of the gas through the nozzle leads to a significant reduction in both its temperature and pressure, thus reducing the Doppler and collisional broadening of the spectral lines compared to when the laser ionization is performed in the gas cell \cite{Kudryavtsev2013}. Consequently, the laser spectroscopy measurements in the jet can reach a spectral resolution as low as 200~MHz FWHM for (super-)heavy elements. With this gas-jet development, the in-gas laser ionization and spectroscopy (IGLIS) technique has been proposed as a powerful tool to study the static electromagnetic properties of radioactive nuclei, such as mean-square charge radii and for many cases spins, magnetic and quadrupole moments. Although not achieving the spectral resolution of collinear laser spectroscopy, this technique offers an excellent compromise between resolution and efficiency, preferred for the most exotic cases \cite{Yang_2023}. After a proof-of-principle experiment at the LISOL facility in Louvain-La-Neuve \cite{Ferrer2017}, the technique has been implemented at several facilities worldwide \cite{Sonoda2013,Hirayama2017,Raeder2020,Zadvornaya2023}, including  S$^3$-LEB \cite{Ajayakumar2023}. Recently, a first measurement of in-gas-jet laser spectroscopy of nobelium was performed at GSI-FAIR using the JETRIS setup \cite{Lantis2024}.    

Currently, the S$^3$-LEB setup has been constructed and is installed at the S$^3$ focal plane, in preparation for first experiments with radioactive ion beams \cite{Ajayakumar2023}. While the current gas cell is designed to enable a significant part of the experimental program \cite{Ferrer2013,Kudryavtsev2016}, future designs are being considered for the study of nuclei with very short half lives or very low production rates, where extraction time and neutralization efficiency become limiting factors. The prototyping and iterative improvement of these designs requires an off-line experimental  setup that reproduces the operational conditions of S$^3$-LEB, but provides simplicity, easy access and is independent of the schedule of experimental campaigns with radioactive beams from S$^{3}$. In this work, we present the design, simulation and first tests performed with such an experimental setup constructed in the framework of the Fast Radioactive Ion Extraction and Neutralization Device for S$^3$ (FRIENDS$^3$) project \cite{FRIENDS3,Dong2024}, which aims to develop a novel gas cell for S$^3$-LEB. 

The article is organized as follows. Section~\ref{motivation} briefly presents the motivation of the FRIENDS$^{3}$ project which has guided the design of the experimental setup and the main requirements that it must satisfy. Section~\ref{design} describes in detail the resulting design, for which ion-transport simulations are presented in Section~\ref{simu}. Finally, Sections~\ref{hardware} and \ref{tests} present the hardware choices for the experimental setup and the results of a first test of its performance using stable alkali ions, including a comparison to simulation results.  

\section{FRIENDS$^{3}$ project}
\label{motivation}

The performance of the S$^3$-LEB gas cell depends in a critical way on three parameters, namely the stopping/extraction efficiency, the extraction time and the neutralization efficiency of the radioactive species of interest. The neutralization efficiency is essential in order to apply the IGLIS technique, which by definition increases by one the charge state of the probed species. The ionization potential of 1+ ions (to the 2+ charge state) is considerably larger than for their neutral counterparts, which often makes the IGLIS technique impractical for ionic species. Consequently, one requires in most cases the neutralization of the incident beam. 

A fraction of the beam is neutralized upon entering the gas cell by charge exchange with the buffer gas \cite{Lautenschlager2016}, but after thermalization charge exchange is no longer energetically possible, so the remaining neutralization must proceed via recombination with electrons removed from buffer-gas atoms in the initial thermalization phase \cite{kudryavtsev2001gas,Facina2004,Moore2010}. The density of these electrons, which directly impacts the recombination lifetime, depends critically on the intensity of the beam entering the gas cell and on the absence of any electrical fields. In LISOL, for example, the very intense cyclotron beam producing the fusion-evaporation reactions traversed the gas cell, creating a permanent electron cloud \cite{kudryavtsev2001gas}. At S$^3$, however, only the reaction products enter the gas cell, which for less intense beams will lead to a limited ionization of the buffer gas \cite{Ferrer2013}. For S$^3$-LEB it is thus interesting to develop a neutralization technique independent of the incident beam intensity, which is the first objective of the FRIENDS$^3$ project.

\begin{figure}
\centering
\includegraphics[width = 0.48\textwidth]{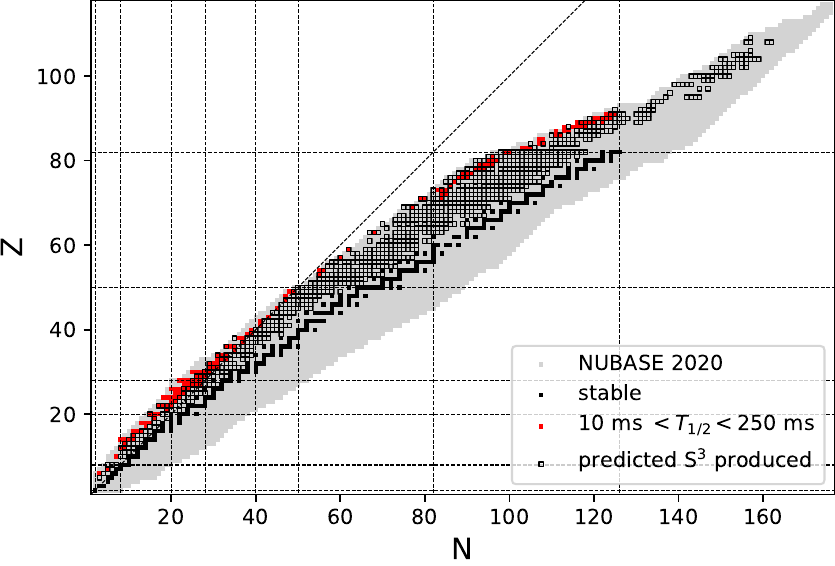}
\caption{Chart showing the nuclides from the NUBASE~2020 database \cite{NUBASE2020} (grey background) and highlighting the isotopes predicted to be produced at S$^3$ (open black squares) or having half-lives between 10 and 250~ms (filled red squares).}
\label{S3-prod}
\end{figure}

The second objective concerns the reduction of the extraction time of the radioactive species, which is about 630~ms for the existing S$^3$-LEB gas cell when using a nozzle with a 1~mm exit-hole diameter \cite{Kudryavtsev2016}. The reason for the long extraction time is the reliance exclusively on gas flow for extraction combined with the very low gas velocity in the ion-stopping volume. This can be penalizing for nuclei or isomeric states with half-lives below 500~ms, as is the case for some of the species predicted to be produced at S$^3$. Figure~\ref{S3-prod} shows a chart based on the NUBASE~2020 evaluation \cite{NUBASE2020}, highlighting the nuclei predicted to be produced at S$^3$ and those with half-lives between 10 and 250~ms in the ground state (10~ms is considered a feasibility limit). One notices that some of the most exotic nuclei expected to be produced at S$^3$ are in this half-life range, especially the $N \approx Z$ nuclei and the very neutron-deficient refractory elements and actinides. Although the extraction time can be reduced to some extent by increasing the diameter of the exit hole (at the cost of required gas consumption and pumping capacity), moving into the 100~ms range can only be achieved by boosting the ion extraction with an electrical field. Such a gas-cell design cannot however rely on neutralization by recombination with electrons in the stopping volume, because the electrical field would remove any electrons before recombination could take place. 

A solution combining the fast transport of the stopped radioactive species and allowing their extraction from the gas-cell in neutral form would be a hybrid design in which an electrical field is used to accelerate the extraction of the ions from the stopping area (where the gas flow is very slow), their transfer to a region of pure gas flow near the exit of the gas cell (without any electrical field) and the application of a neutralization mechanism in this volume. Following neutralization, the species would be carried out of the gas cell by the flow alone in a few tens of milliseconds. Building on the approach which was successfully applied in the RADRIS technique \cite{Lautenschlager2016, Laatiaoui16}, a gas-cell design was proposed and tested on-line at GSI combining extraction by electric field with collection on a heated filament placed in a channel close to the gas-cell exit \cite{Raeder2020, Lantis2024}. Ions collected on the filament would be neutralized and evaporated in atomic form in the gas flow towards the gas-cell exit. 

In the FRIENDS$^3$ project we aim for a more universal method, thus eliminating the need of collecting the ions on a filament and instead neutralize them during their flow through the channel, before extraction from the gas cell. A critical point for this type of extraction is the transfer of ions between the areas with and without electrical field, due to the tendency of the ions to follow the field lines to their termination on the gas-cell walls, where they are lost. To minimize losses, the electrical field lines must be aligned with the gas flow as well as possible at the entrance of the no-field region. Furthermore, it is beneficial if the gas velocity in this region is significant, in order for the viscous force to guide the ions towards the gas-cell exit. Detailed simulations of ion transport in gas were used to design a gas-cell of this type \cite{Dong2024} and are presented in a separate article \cite{Dong2025}. Here, we focus on the description of the experimental setup as a whole. 

The setup of the FRIENDS$^3$ project is required to recreate the experimental conditions of an IGLIS experiment. 
It should allow implementing a gas cell, producing a supersonic gas jet and coupling the required laser beams for ionization either in the gas cell or gas jet. 
Sufficient space and easy access should be ensured in the jet chamber for laser alignment. 
The photo-ions should be guided out of the gas-jet area with high efficiency and delivered to a high-vacuum chamber where they can be individually counted. 
This requires implementing differential pumping stages and ensuring sufficient pumping throughput. 
A particular condition is versatility with respect to the exit geometry of the gas cell, allowing if needed to implement and test different designs.  

A pressure in the gas-jet chamber of the order $10^{-1}-10^{-2}$ mbar is necessary for a jet Mach number between 5 and 10 for a typical gas-cell stagnation pressure of 100~mbar (see e.g. Fig.~11 in \cite{Kudryavtsev2013}).
Considering that the operation of a microchannel plate (MCP) detector for single-ion counting requires a pressure on the order of $10^{-6}$~mbar, the setup would require an intermediate differential pumping stage on the level of $10^{-4}-10^{-3}$~mbar. 
Ion transport in these pressure conditions between small apertures can only be achieved by guiding the ions in RFQs. 
The transport line should also contain a mass-separation stage, allowing at least coarse purification of the photo-ions from contaminants. 
An RFQ operated in quadrupole mass filter mode is required for this function. 

The resulting design, which will be detailed in the next section, bears similarities to the IGLIS setup at LISOL in Louvain-la-Neuve \cite{Kudryavtsev2013, Ferrer2017} and the JetRIS setup at GSI in Darmstadt \cite{Raeder2020,Lantis2024}.

\section{Description of the setup}
\label{design}

Following simulation work, which will be presented in the next section, a setup satisfying the requirements mentioned above was designed and is shown in the 3D drawing of Fig.~\ref{3D-setup}. 
The drawing is presented in top-down, half-section view with respect to the horizontal plane containing the axes of the gas cell and of the three RFQs. 
The support structure and the vacuum pumps are visible behind the vacuum chambers. 
In the following, the different components of the setup will be described.


\subsection{Gas-cell}

\begin{figure}[!ht]
\centering
\includegraphics[width=0.48\textwidth]
{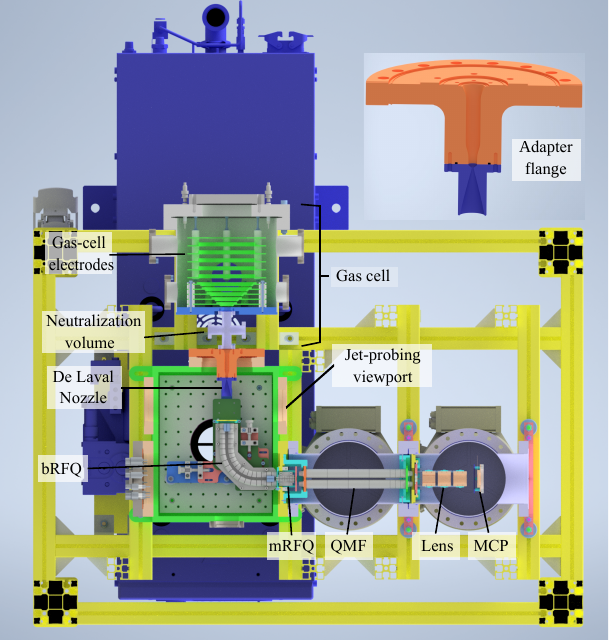}
\caption{3D drawing of the experimental setup in half-section view with respect to the horizontal plane containing the gas-cell axis. The upper-right inset shows a magnified view of the adapter flange and de Laval nozzle. For details, see text.}
\label{3D-setup}
\end{figure}

The electrical gas-cell, visible in the top part of Fig.~\ref{3D-setup}, consists of a cylindrically-symmetric electrode structure placed in a ultra-high vacuum (UHV) DN250 chamber with conflat (CF) flanges. 
The choice of the chamber size, electrode diameter and length is a compromise between, on the one hand, the requirement of stopping the fusion-evaporation products from S$^3$ and ensuring a minimal distance between the electrodes and the surrounding vacuum chamber (to avoid sparks) and, on the other hand, the possibility to install the gas cell in the existing front-end chamber of the S$^3$-LEB setup (to simplify implementation in case of future on-line tests). 

The cage is 72~mm long with an internal diameter equal to 110~mm, consisting of 5 electrodes. Each electrode is 8 mm long. 
The cage electrodes are separated from one another by a distance of 8 mm with alumina insulators surrounding the fixation rods. They are also separated from the entrance flange (top flange in Fig.~\ref{3D-setup})  by a ring electrode with the same outer diameter as the cage, but an inner diameter of 80~mm. This opening is sufficient to install an entrance window similar to the one designed for S$^3$-LEB (50~mm window diameter and 74~mm mounting flange diameter). The entrance window, not represented in Fig.~\ref{3D-setup}, would be installed on the entrance flange. 
The five funnel electrodes have decreasing inner diameters from 110 mm to 16 mm. Each electrode is also 8 mm long and all the electrodes are separated and insulated from the others by a longitudinal distance of 2 mm using alumina insulators as well. 
The cage is fixed on the entrance adapter flange whereas the funnel electrodes are fixed on the exit plate.
The centering is mechanically constrained by the fixation of both sets of electrodes on their respective supporting surface. 

The DN250 chamber containing the gas cell electrodes is followed by a CF DN16 cross used as a neutralization channel. 
The gas cell is connected to the rest of the setup by an adapter flange which is represented in Fig.~\ref{3D-setup}. 
A tube with a smooth inner contour is built within the flange, ensuring the final convergent part of the gas cell from a 16~mm diameter~at the beginning to a 6~mm diameter at the end. 
The entrance face has CF16, CF40 and CF100 grooves, allowing to seal the three types of flanges on it. The exit face allows installing a de Laval nozzle, which is also represented in Fig.~\ref{3D-setup}. 
The nozzle is designed by KU Leuven and produces a 60 mm-long collimated gas jet of Mach number 8 \cite{Ferrer2021,Claessens2024}. The gas cell can be evacuated by an Edwards nXR40i primary pump.


\subsection{Ion extraction, guiding and detection}
\label{transport}

The ion transfer section following the gas jet  consists of the three RFQs and an Einzel lens to focus the ions on an MCP. To ensure good vacuum conditions for its operation, differential pumping is applied, leading to the use of three different chambers separated by apertures. 

The first chamber is non-standard, of rectangular shape, containing the de Laval nozzle as well as the first RFQ (bRFQ in Fig.~\ref{3D-setup}). This is the chamber were the gas jet is formed. Two CF63 viewports laterally aligned with the gas jet allow overlapping an expanded laser beam for gas-jet ionization. A third CF40 viewport allows longitudinal access for a laser traversing the bRFQ in counter-propagating direction. The ions of interest are then extracted from the jet and transported in the bRFQ through an injection plate. 

The bRFQ is bent in order to allow the counter-propagating laser path and at the same time to avoid injecting gas directly in the mRFQ, improving pressure suppression between the differential pumping regions. It is made of 18 segments, of which five compose the straight part at the entrance, nine in the curvature, and four in the straight part at the exit. The electrodes are fixed between plastic insulator boards which have the role of mechanical support and of basis for a printed circuit implementing a voltage divider chain for the DC voltages applied to the segments. They also have the role of insulating the DC and RF inputs. Its dimensions are 125 mm x 113.5 mm x 29 mm, with an inner characteristic radius of 6 mm and a total length of the longitudinal path of the ions of 189 mm.

The bRFQ is placed on a alignment system made with three plates of which the top-most two allow relative translation. 
This allows to align the bRFQ with the gas-jet direction and the rest of the line in the horizontal plane, thanks to micrometric screws. The vertical position and tilt are controlled with three screws which support the translation system. An optical plate is installed in the chamber under the RFQ, allowing to add optical elements for laser alignment. 


The rectangular chamber is pumped by a GXS160/1750 screw pump, separated from the chamber by a gate valve and a bellows used to decouple the vibrations from the pump. The typical pumping speed of the screw pump is 1160 m$^3/$h. 

The bRFQ is followed by a mini-extraction RFQ (mRFQ in Fig.~\ref{3D-setup}) with the role of transferring the ions into the second chamber. The mRFQ is thus placed at the entrance of the second chamber, behind a PEEK aperture plate separating the two vacuum regions, and mounted on a custom-made CF100 flange. The aperture in the PEEK plate is cut out to surround the first mRFQ segments, which penetrate the volume of the bRFQ. This minimizes the gas transfer from the first to the second chamber. The mRFQ is 36.5 mm long and made of 8 segments of equal length and an inner characteristic radius of 2 mm.
As for the bRFQ, the segments are fixed between two insulator boards which also support the printed circuits for applying the voltages to the RFQ. A second printed-circuit board which is detached from the RFQ implements a voltage  divider chain for the DC voltages and provides the insulation between the DC and RF inputs.  

The first segments of the mRFQ are overlaid by 1~mm with the last segments of the bRFQ to ensure a good transfer with minimal ion losses between the first and second chamber. An extraction plate with 4 mm inner diameter is placed right after the mRFQ, separating it from the following ion guide. Both the bRFQ and mRFQ were constructed and first commissioned at KU Leuven \cite{Ferrer2013}. 

The third RFQ (QMF in Fig.~\ref{3D-setup}) is also installed in the second chamber, after the mRFQ. Its role is on the one hand, to ensure the transport of the ions along the entire length of the chamber (determined by the turbo pump diameter - see below), and on the other hand to separate the ions according to their mass and to allow their identification (playing the role of quadrupole mass filter). It can also eliminate any ionized gas extracted from the gas cell, or other contaminants. The QMF is 199 mm long with 7 segments of two different sizes : 4 segments of 41.5 mm each and 3 segments of 10 mm, all separated by 0.5 mm. In addition, it contains an entrance and exit plate with apertures of diameter 10 mm. 
Its inner characteristic radius is 6 mm. The segments are hollow and are separated from each other and from the entrance/exit plate by insulating rings. The segments and the rings are traversed by stainless steel M6 rods which support their weight and are locked with nuts on the entrance and exit plates, pressing the structure together. The insulating rings have smaller inner diameter than the segments, which avoids charge-up and electrical contact. The rods and nuts being metallic, one need to bias them to avoid the creation of floating elements. The simplest way is to bias them through the nuts fixing them either on the entrance or the exit plate of the QMF (while for the opposite plate the nuts would be insulated through a PEEK washer). From the simulation work presented in section \ref{simu}, the entrance of the ion guide proved to be where the transmission of the ions is more delicate.
Consequently, it was chosen to bias the rods and nuts to the potential of the entrance plate and insulate them from the exit plate.
 
\begin{figure}[!ht]
    \centering
    \includegraphics[width=0.48\textwidth]{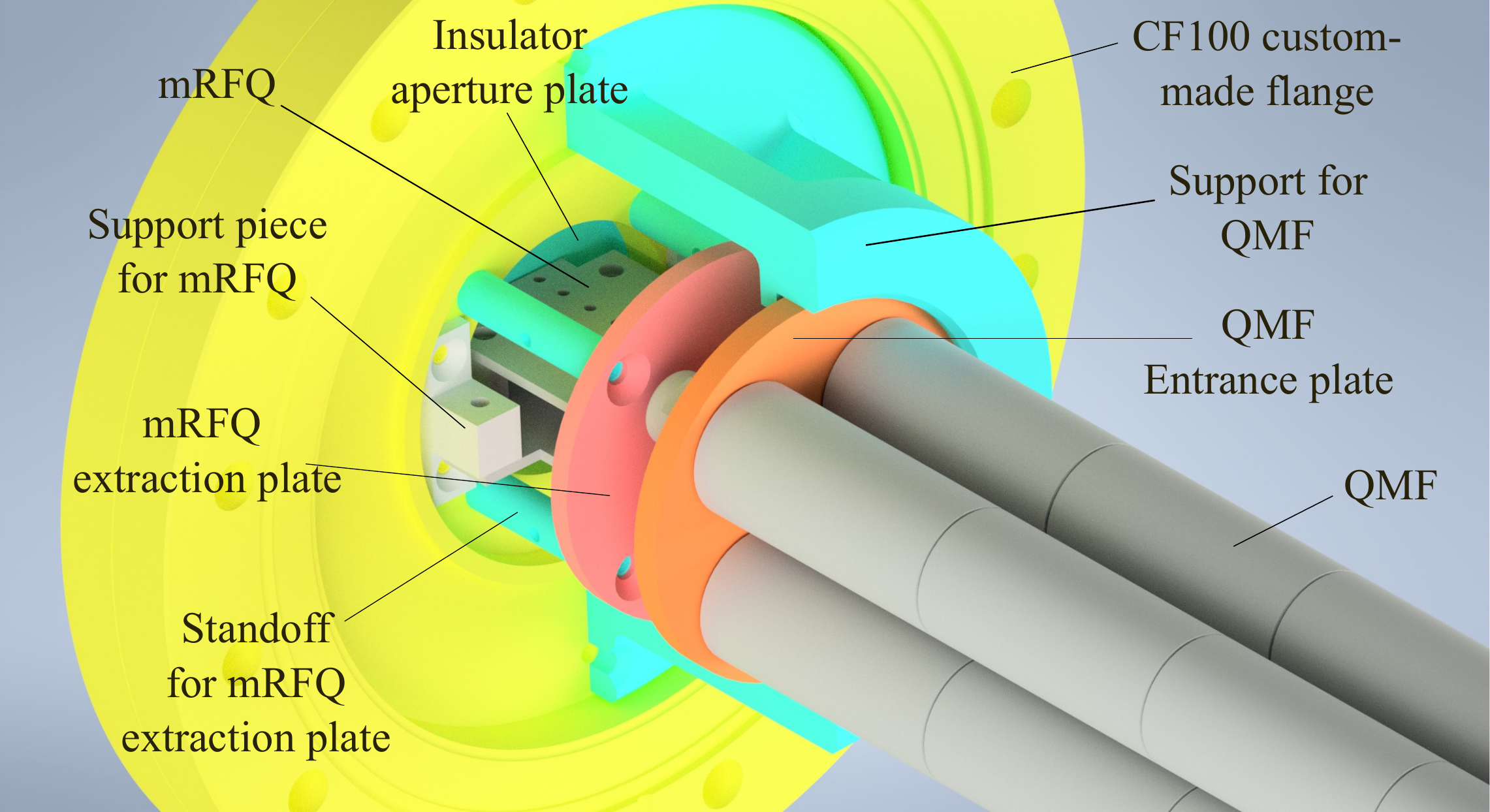}
    \includegraphics[width=0.48\textwidth]{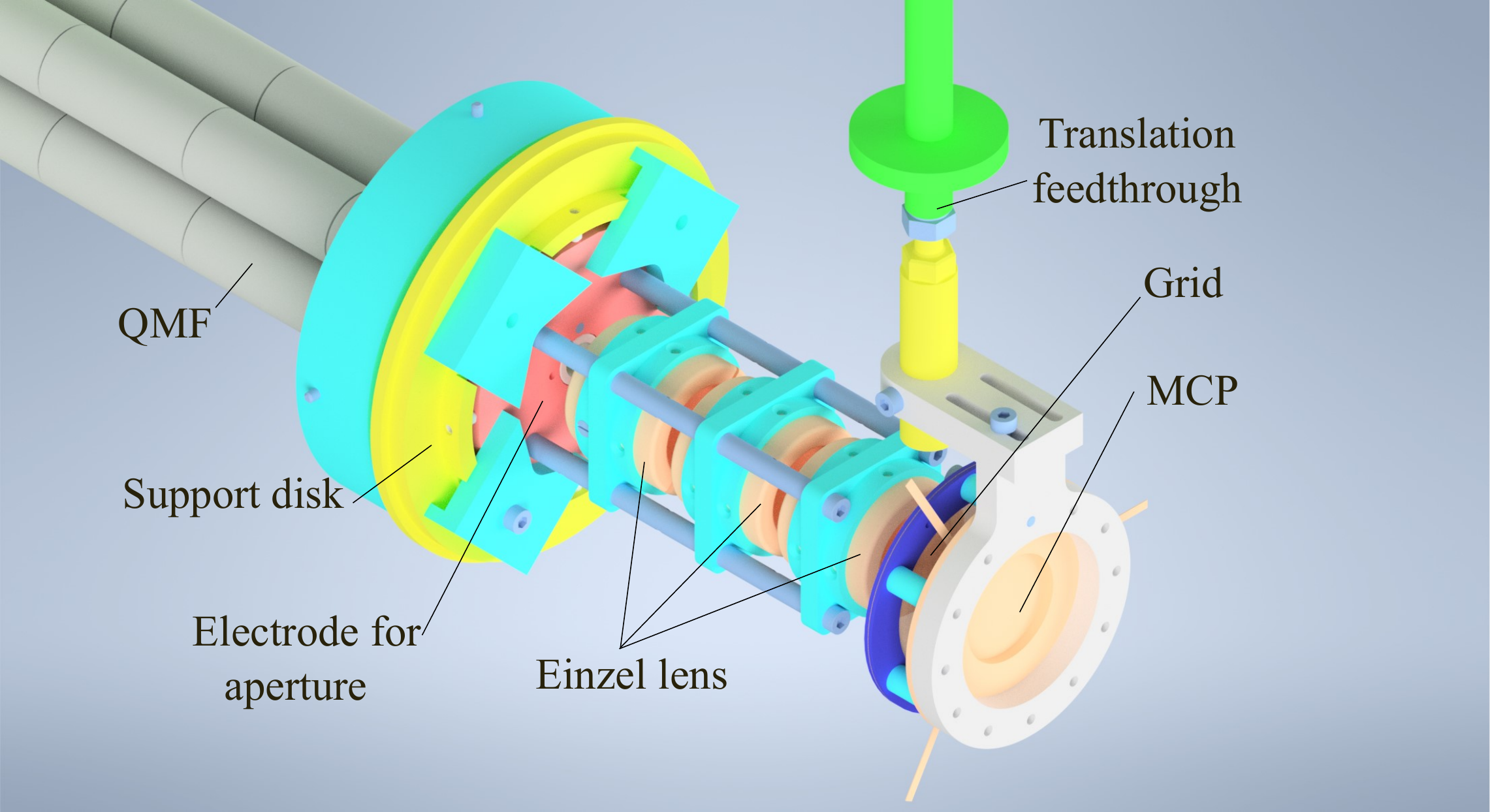}
    \caption{Top : Interface between the bRFQ chamber and the QMF chamber with the mRFQ and the support of the QMF. The PEEK insulator elements are colored in cyan. The QMF support is cut in the view to see the elements inside. The yellow flange is the custom-made CF100 flange, supporting the insulating plate with the aperture for differential pumping, the mRFQ, the extraction plate after the mRFQ (in pink) and the support for the QMF. Bottom : Interface between the QMF chamber and the detection chamber. The PEEK insulator elements are colored in cyan. The yellow circular electrode is a support disk welded to the entrance flange of the chamber, while in pink one observes the supporting electrode for the differential-pumping aperture piece and for the Einzel lens. This electrode is insulated from the support disk by a gasket and nylon-tipped screws. After the Einzel lens shown at the center of the figure, the MCP stack mounted on a translation feedthrough is visible. The vacuum chambers are not represented in the two figures}
    \label{det}
\end{figure}

The QMF is supported from the entrance side by a hollow cylindrical insulator piece made out of PEEK and open on both sides, surrounding the mRFQ (cyan in Fig.~\ref{det}, top, and shown in section). The supporting piece is fixed onto the special CF100 flange separating the first and second chambers (yellow in Fig.~\ref{det}, top, which also holds the mRFQ and the PEEK aperture plate) and is concentric with the optical axis of the two RFQs. Towards the QMF, the piece has a circular slot of the same diameter as the QMF entrance plate (orange in Fig.~\ref{det}, top), allowing it to slide in tightly and be centered with respect to the mRFQ axis. The exit side of the QMF is fixed on a cylindrical PEEK insulator (left-most cyan piece in Fig.~\ref{det}, bottom) with a diameter slightly smaller than the chamber (100 mm). Three radial screws traversing the wall of the piece  towards the exterior allow it to be centered and fixed to the chamber. 

The second chamber (designated as QMF chamber) as well as the third one (designated as detection chamber) are standard CF160 to CF100 reducing crosses, with the DN 160 axis orientated vertically. The interface between the QMF chamber and the detector chamber, is a grounded plate (pink in Fig.~\ref{det}, bottom) with a changeable aperture piece allowing to modify the diameter of the aperture to either 8 mm, 6 mm or 4 mm. 

Going to lower diameter improves the vacuum in the detection chamber, which would be required in case the buffer-gas pressure needs to be increased, but degrades the transmission to the detection area. The aperture plate is pressed by four rectangular wings onto a ring welded in the entrance flange of the chamber (yellow in Fig.~\ref{det}, bottom), insulated by a gasket. It is centered by using four nylon-tipped screws traversing the sides of the ring.

The Einzel lens, visible in Fig.~\ref{det}, is fixed on the aperture plate with metallic struts (thus floated to the same potential). It consists of three electrodes of equal size (26 mm long, 20 mm inner diameter) with 5 mm spacing. The first electrode is cut cross-wise into four to make a steerer. The MCP detector is placed 15 mm downstream from the Einzel lens. It is fixed on a translation feedthrough allowing its vertical alignment. 
A grid is installed in front of the MCP to reduce the transmission to 10 $\%$ and measure the beam current when higher intensities are used.

The QMF and detection chamber are each pumped from below by their own Pfeiffer HiPace700 turbo pump. The primary vacuum is achieved by a common Edwards nXR120i primary pump, providing a typical pumping speed of 120 m$^3/$h.

\section{Simulation studies}
\label{simu}
The design of the experimental setup was simulated in order to optimize the electrical potentials, to quantify the expected performance in the planned operation modes, as well as to test the effect of certain geometry choices. The simulations were performed using the SIMION$^{\copyright}$ software \cite{Dahl2000}.

Residual gas is implemented in each region of the simulation and a collision model is used to compute ion trajectories taking into account the scattering between ions and the gas atoms. The collision model used in the present is the so-called hs1 model of SIMION$\copyright$ \cite{Manura2007}.
The ions travel in a neutral gas, elastically colliding and scattering with the atoms according to the hard sphere model of kinetic theory.
The ion and the gas atom are considered to be interacting in a cylindrical volume.
The corresponding area of the cylinder is  defined by the interaction cross section which is that of a circle having as radius the sum of the kinetic radii of the two colliding particles such that $\sigma = \pi (r_{ion} + r_{gas})^{2}$ with $r_{ion} = r_{Cs} = 1.85 \times 10^{-10}$ m and  $r_{gas} = r_{Ar} = 1.83 \times 10^{-10}$ m.
The resulting cross section for the interaction between the cesium ions and the argon buffer gas atoms is 4.25$\times 10^{-19}$ m$^{2}$.

The collision frequency $Z$ is the integral of the collision frequencies for each possible speed  $c$ of the gas particle, leading finally to $Z = \sigma n c$ with $n$ being the density of gas atoms.
$Z$ is related to the average of the velocities weighted by their population densities.
The mean-free path $\lambda$ of the ions traveling and colliding with the gas atoms is therefore $\lambda = c_{ion}/Z = c_{ion}/(\sigma n c)$.

\begin{figure}[!ht]
     \centering

\includegraphics[width=0.47\textwidth,trim={0.2cm 0 0 0},clip]{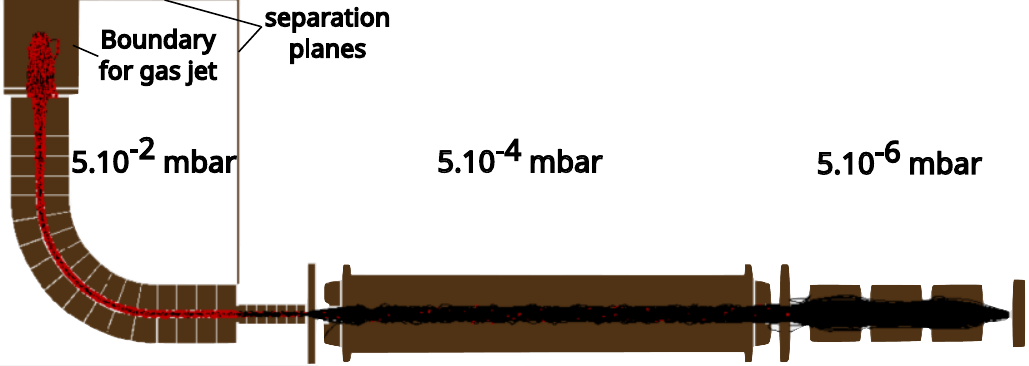}
         \caption{2D view of the 3D ion optics simulated with SIMION$^{\copyright}$ for the transport in the experimental setup. Pressures indicated are for argon buffer gas.}
        \label{Simulation}
 \end{figure}

In the present simulation work, the mean free path is automatically  calculated from the pressures (shown in Fig.~\ref{Simulation}) and temperature (taken as 300 K).
In each time step, the collision probability is calculated from the mean free path. 
From this probability, the software determines whether a collision took place or not.
The trajectory calculation is then performed assuming an elastic collision. 

The simulated transport line contains the three segmented RFQs and the Einzel lens used to focus the ion beam on the MCP detector.
The simulation volume is shown in Fig.~\ref{Simulation}.
The grid scaling of the simulation is 2 grid units/mm and the geometry has no symmetry. 
It is divided in three areas with different pressures.
The first area contains the gas jet and the bRFQ. 
Its dimensions are 45 mm x 46 mm x 34 mm.
A box at ground potential was added in the simulation around the jet in order to create proper boundary conditions for the relevant potentials. 
To properly define the boundary conditions for the whole chamber, two additional separation planes are  added as shown in Fig.~\ref{Simulation}: one (perpendicular to the gas jet), marking the beginning of the bRFQ chamber along the gas-jet axis, and a second one, perpendicular to the bRFQ exit axis, marking the separation between the bRFQ and mRFQ chambers. 
The pressure in the first area is set to 5$\times10^{-2}$~mbar.

The separation between the first and second volumes of the simulation, corresponding in reality to the PEEK insulator between the bRFQ and mRFQ, is simulated by a  test of the displacement from the RFQ center axis of the ions exiting the bRFQ. Any ion crossing the exit plane at a distance larger than the radius of the aperture is considered lost and removed from the simulation. 
This virtual conductance has a 9.5 mm inner diameter centered on the RFQ axis. 


The second simulation volume contains the mRFQ and the QMF.
Two pressures were typically used in the simulations, either $5 \times10^{-4}$ or $10^{-3}$~mbar.
Finally, the third volume contains the Einzel lens and the MCP detector and the used pressure is $5\times10^{-6}$ mbar. The pressures were estimated using the throughput equation and considering the conductances and pumping speeds.
All the simulation work presented in the following was performed with $^{133}$Cs$^{+}$ ions, because they were also readily available from a surface ion source for testing the setup. 
The ions are created in the simulation at 
the middle of the gas jet, being Gaussian distributed in the transverse plane such that the standard deviation of the distribution is equal to 1 mm.
Their speed follows a Gaussian distribution as well, with a mean value equal to the one they are supposed to have in the jet i.e. 550~m/s and a standard deviation of 50~m/s.
Their initial distribution of velocity directions is defined as a cone of half angle equal to 15~$\deg$. We note that these initial conditions of the simulated ion beam are worse than the ones provided experimentally, as the jet divergence is typically much smaller. 

The potentials of all the electrodes were optimized to maximize transmission, prioritizing when possible lower potentials and RF amplitudes. 
Generally, the transport is ensured by DC potentials, which have to be negative (because the gas cell is on ground potential) and gradually become more negative  as the ions advance along the setup. Potential gradients are needed in the bRFQ and mRFQ to optimize ion guiding and avoid trapping the ions after collisions with the buffer gas. However, excessive gradients result in unstable ion trajectories following collisions.  

Following optimization, a set of reasonable settings, given in Table~\ref{transmset}, allow obtaining up to 88 $\pm$ 2 \% total transmission. The uncertainties are calculated by running the same simulation 5 times with resampled ion distributions and computing the statistical rms deviation. The choice of RF frequency for the optimization was 1~MHz, which is a typical value for ion guides, allowing good RF guiding without requiring too large RF power.
The main transmission losses occur according to simulations in the QMF area. Throughout the article, the $\pm$ values show the minimal/maximal dispersion based on several runs with 1000 ions, each time resampling the initial distributions.  
Almost no ions are lost in the mRFQ, the total transmission through the bRFQ and the mRFQ being 98 $\pm$ 1 \%, with the small losses occurring at the bRFQ entrance plate. 
Of the initially produced ions, 94 $\pm$ 1 \% are transferred through the QMF and 92 $\pm$ 1 \% pass through the final aperture towards the MCP.
In the transport between the QMF and the detector, the ions are lost either in the collimator or in the Einzel lens, the latter loss concerning ions which have the largest scattering angles on the rest gas in the QMF chamber.
A compromise in the voltages has been made for the collimator and Einzel lens potentials to maximize the transmission of the whole.


\begin{table}[t]
\centering
\begin{tabular}{c l c c}
 \multirow{2}{*}{Section} & \multirow{2}{*}{Element} & \multicolumn{2}{c}{Voltage (V)} \\ &  & 1 MHz & 500 kHz \\ 
 \hline
 \multirow{4}{*}{bRFQ} &  RF amplitude  & 175  &  30  \\
  & entrance plate & -5 & -4 \\
  & DC in & -11 & -10.5 \\
  & DC out & -28 & -19 \\
  \hline
   \multirow{4}{*}{mRFQ} & RF amplitude & 25 & 8 \\
  & DC in & -28.2 & -19.1 \\
  & DC out & -29.4 & 19.7 \\
  & exit plate & -30 & -20 \\
   \hline
  \multirow{5}{*}{QMF} & RF amplitude & 175 & 38 \\
  & entrance plate & -65 & -50 \\
  & DC in & -52 & -36 \\
  & DC out & -64 & -42 \\
  & exit plate & -72 & -57 \\
   \hline
     \multirow{4}{*}{Detection} & conductance & -400 & -400 \\
  & Lens el. 1,3 & -175 & -150 \\
  & Lens el. 2 & -400 & -375 \\
  & MCP & -2000 & -2000 \\
   \hline
\end{tabular}
\caption{Optimal potentials applied to the whole transport line in both 1~MHz and 500~kHz configurations}
\label{transmset}
\end{table}

The RF amplitudes and gradient strengths do not impose limitations on the transmission.
For example, the transmission may vary by only a few percent with a change by a few tens of Volts in the RF amplitudes in the bRFQ and the ion guide. 
A more critical aspect of the optimization is the matching between the ions' emittance at the exit of one element (especially an RFQ) and the acceptance of the following one, or the transfer between two elements separated by a drift zone, as the ion trajectories tend to diverge once no longer confined by the RF field. 


As mentioned in Section~\ref{transport}, the possibility of changing the diameter of the aperture separating the QMF and the detection chambers was integrated in the design of the transport line.
Simulations were run to quantify the losses induced by the reduction of the aperture diameter to 6~mm and 4~mm.
The results are given in the Table~\ref{cond}.

\begin{table}[!ht]
\centering
\begin{tabular}{lccc}
\hline
Aperture diameter (mm) & 8  & 6  & 4 \\
\hline
Transmission ($\%$) & 88 $\pm$ 2 & 82 $\pm$ 3 & 65 $\pm$ 3 \\
\hline
\end{tabular}
\caption{Simulated transmission efficiency as a function of the diameter of the aperture between the QMF and detection chambers, for 1~MHz RF frequency}
\label{cond}
\end{table}

The reduction of the transmission is not linear with respect to the decrease of the aperture diameter. 
Switching to the 6~mm diameter could therefore be a good option to help for the differential pumping, as the difference in transmission with respect to 8~mm is not too important. 
The transmission with the 4~mm diameter aperture is nevertheless still acceptable if one needs to reduce even more the gas conductance between the two chambers.

In addition to the standard study of the ion transport with the RFQs in the optimal positions and conditions chosen for the design, studies were performed to investigate the impact of certain changes in the setup geometry or operation mode. These are the impact of an axial shift of the mRFQ with respect to the bRFQ, the choice in biasing the QMF support rods and nuts, the use of a lower RF frequency, as well as the use of the QMF to bunch the beam. 


The rigidity of the mRFQ supports and the cumulation of machining errors raised the possibility that the final axial position of the mRFQ would differ from the expected one by up to 2~mm. 
Simulations were thus run to quantify the impact of such a shift in the position of the mRFQ with respect to the bRFQ. 
The results are summarized in Table~\ref{shift}, where the shift of the mRFQ is considered in the direction of the QMF. The simulations do not show an effect of the shift of the mRFQ within the considered range.

\begin{table}[!ht]
\centering
\begin{tabular}{lccc}
\hline
Shift (mm) & 0 & 1 & 2 \\
\hline
Transmission ($\%$) & 88 $\pm$ 2 & 89 $\pm$ 1 & 90 $\pm$ 2 \\
\hline
\end{tabular}
\caption{Simulated change in transmission efficiency when shifting the mRFQ out of the bRFQ}
\label{shift}
\end{table}



As discussed in section \ref{transport}, the fact that the rods and nuts holding together the segments of the QMF are metallic implies the need to bias them to avoid an unwanted focusing effect or floating elements in the beam line.
We performed simulations to check whether it is more favorable to bias them (by direct contact) at the potential of the QMF entrance or exit plate.
The results 
show no difference between the two configurations.

 
For the basic optimization of the setup parameters, an RF frequency of 1~MHz was considered for all RFQs. Nevertheless, this choice can lead to rather large RF amplitudes needed to transfer the ions especially in mass filtering mode and for ions of masses towards or above $A\approx 200$. For the latter cases, it is also potentially interesting to operate the mass filter at 500~kHz, which according to the formula of the Mathieu $q$ stability parameter of the QMF would require a factor 4 lower amplitude for achieving optimal filtering at $q$~=~0.706 \cite{Major2005}. This can also be beneficial for the bRFQ, for which the use of lower RF amplitude reduces the risk of discharges.  

All the simulations were therefore performed also for an RF frequency of 500~kHz (applied to all RFQs). We found that the optimal transport potentials depend on the RF frequency. Table~\ref{transmset} shows the reoptimized potentials. To maximize transmission at 500~kHz, it was necessary in the simulation to reduce the axial gradient in the RFQs by a factor of 2. Especially in the bRFQ, this proportionally reduced the ion drift velocity and ensured that the number of RF cycles seen by the ions during the RFQ transit time was kept the same, thus ensuring a similar radial cooling as for 1~MHz RF frequency.  


The 500~kHz transport settings led to the simulated transmission efficiencies presented in Table \ref{cond500k} for the different diameters  of the aperture between the QMF and the detector chamber. One observes a slightly lower transmission efficiency in ion guiding mode at 500~kHz than 1~MHz. 

 

\begin{table}[!ht]
\centering
\begin{tabular}{lccc}
\hline
Aperture diameter (mm) & 8  & 6  & 4 \\
\hline
Transmission ($\%$) & 75 $\pm$ 2 & 75 $\pm$ 3 & 63 $\pm$ 3 \\
\hline
\end{tabular}
\caption{Simulated transmission efficiency as a function of the diameter of the aperture between the QMF and detection chambers, for 500~kHz RF frequency}
\label{cond500k}
\end{table}

In addition to the optimization test made in transmission mode, the ion guide has been configured either as a mass filter or as a buncher. The mass filtering mode of the setup was simulated for both 500~kHz and 1~MHz configurations. In the mass filtering mode of an RFQ, one adds to the constant DC offset $V_0$ of each given segment a quadrupole voltage distribution of amplitude $V_{DC}$, such that one pair of opposite segments is biased to $V_0 + V_{DC}$ and the other pair is biased to $V_0 - V_{DC}$. Considering also the RF amplitude $V_{RF}$, this will lead to a characteristic pair of Mathieu parameters $a$ and $q$ given by \cite{Major2005}:

\begin{equation}
    a = \frac{8 e V_{DC}}{m r_0^2  \omega^2}, q = \frac{4 e V_{RF}}{m r_0^2  \omega^2},
    \label{Mathieu}
\end{equation}
where $e$ and $m$ are the ion charge and mass, respectively, $r_0$ is the inner RFQ radius and $\omega$ is the angular frequency of the RF driving field. To perform a so-called mass scan with an RFQ configured in mass-filter mode, one scans the applied voltages with a constant ratio $V_{DC}/V_{RF}$  to change the range of masses transmitted, with the resolving power increasing as the scan line is moved close to the upper tip of the first stability region with coordinates $a < 0.236$ and $q = 0.706$ (which is the position of the peak of the first stability region of the device \cite{Major2005}).
Then, one can systematically vary the set mass $m$ and apply the $V_{DC}$ and $V_{RF}$ values which ensure for the given mass the chosen $a$ and $q$ Mathieu parameters. Recording the device transmission as a function of the set mass-to-charge ratio one obtains a mass spectrum. Higher resolving power typically leads to lower transmission. 

Here, we have simulated the mass filtering mode both for the QMF and the mRFQ and the two tested frequencies, 500~kHz and 1~MHz. 
Table~\ref{filter} summarizes the transport efficiencies obtained for different values of $a$ close to the theoretical maximum. For each result, a mass scan is simulated for a beam consisting of only $^{133}$Cs$^+$ ions. The resulting transmission curve is fitted with a Gaussian and its FWHM $\Delta m$ is determined. The resolving power $R = m/\Delta m $ is then calculated. The quoted efficiency is for the maximum of the transmission peak. 

\begin{table}[!ht]
\centering
\begin{tabular}{lcccc}
\hline
RFQ & RF frequency & $a$ & $R = m / \Delta m$ & efficiency \\
\hline
\multirow{2}{*}{mRFQ} & 500 kHz & 0.22 & 16 & 4 $\%$ \\
 & 1 MHz & 0.23 & 37 & 5 $\%$ \\
\hline
\multirow{4}{*}{QMF} & \multirow{2}{*}{500 kHz} & 0.225 & 17 & 16 $\%$ \\
 &  & 0.23 & 25 & 5 $\%$ \\
 \cline{2-5}
 & \multirow{2}{*}{1 MHz}  & 0.225 & 21 & 25 $\%$ \\
 &  & 0.23 & 31 & 14 $\%$ \\
\hline
\end{tabular}
\caption{Simulated performances of the mRFQ and QMF in mass filter mode.}
\label{filter}
\end{table}

One observes that, although the mRFQ can be used as a mass filter, its efficiency is significantly lower (about a factor 3 for the same value of Mathieu $a$). Furthermore, using a lower RF frequency reduces both the efficiency and the resolving power of the device (be it mRFQ or QMF). 

For future applications, we have also simulated the possible use of the QMF as a buncher without any additional gas injection. For pressures in the gas-jet chamber on the order of $10^{-1}$~mbar, one expects the QMF to operate in the few $10^{-3}$~mbar regime, which could be sufficient for ion trapping. As discussed in Section~\ref{design}, the last 3 segments of the QMF are only 10~mm long, which allows creating a narrow trapping region. We therefore optimized the QMF potential for trapping the ion beam and simulated the resulting bunch width on the MCP, for the two RF frequencies that were investigated. For these simulations, the Mathieu $a$ parameter of the QMF was set to 0. 
Table~\ref{buncher} shows the obtained results. They were obtained by changing the potentials in the QMF, fixing the gradient to 1 V per electrode to lower the energy of the ions in the QMF. When applying RF frequency of 500 kHz, the energy of the ions is too high to trap them correctly. This can either be compensated by decreasing the gradients in all the RFQs and in particular in the mRFQ and the bRFQ or by increasing the amplitude in the QMF. The results in Table~\ref{buncher} are obtained by fixing the RF amplitude to 80 V (zero to peak) in the QMF. In addition, the voltage on segment 7 (last segment) of the QMF has been increased to create a potential barrier and the voltage of segment 6 was lowered to create a potential well for the trapping. To extract the ions, segment 5 and 6 were switched after 5 ms to 130 V and 30 V respectively.
The optimal values depended on the used pressure and RF frequency. The offset potential of the QMF was made less deep to better decelerate the ions entering the trap.
These simulations were performed with the same pressures as the ones in transmission and mass filtering mode and also by increasing the pressure in the QMF chamber to 10$^{-3}$~mbar. 

\begin{table}[!ht]
\centering
\begin{tabular}{lccc}
\hline
Pressure (mbar) & RF frequency  & \shortstack{bunch\\width (ns)} & \shortstack{efficiency} \\
\hline
\multirow{2}{*}{1$\times$10$^{-3}$} & 500 kHz  & 160(5) & 14 $\%$ \\
 & 1 MHz  & 104(2) & 31 $\%$ \\
\hline
\multirow{2}{*}{5$\times$10$^{-4}$} & 500 kHz  & 223(7) & 12 $\%$ \\
 & 1 MHz  & 145(4) & 10 $\%$ \\
\hline
\end{tabular}
\caption{Simulated transmission efficiency and bunch width (FWHM) of the QMF operated in bunching mode. }
\label{buncher}
\end{table}

One observes that the trapping efficiency without additional gas injection in the QMF is low. 
Nevertheless, it is possible to trap at least part of the beam, either for direct in-trap experiments or to produce bunches of a few 100~ns FWHM for injection in other traps downstream.




\section{Hardware and control system}
\label{hardware}

The different DC voltages required to operate the FRIENDS$^3$ experimental setup are provided by a combination of commercial low-voltage and high-voltage power supplies, either with numerical or analogical control. The latter case corresponds to a series of Spellman MS DC-DC modules providing a maximum of -500~V and controlled using an input voltage of 0-5 V. The control voltage is delivered by a dedicated multichannel power supply, either two Moku:Go multifunctional FPGA devices from Liquid Instruments, or a Wago 750, 12 bit analog-output module.   

The RF signals required by the bRFQ and the QMF are delivered by an Aim-TTi TG2512A 2-channel function generator, amplified by two HLA150V-PLUS linear amplifiers from RM Italy. For the mRFQ, which is less demanding in power, the RF signal is generated by one of the Moku:Go devices, being amplified by an Aim-TTi WA301. For each RFQ, an air-core transformer circuit is used, on the one hand, to generate the positive and negative phases of the RF field (by grounding the middle of the secondary coil) and, on the other hand, to impedance match the capacitive load of the RFQ to the amplifier. Amplitudes higher than 500~Vpp were thus obtained for the QMF, which would be sufficient to perform mass scans of ions up to around mass  $A\approx 200$.

For ion detection in high-intensity regime, the ion current was measured using a Keithley picoammeter model 6485. The signal measurement cable was typically connected to the MCP grid, or to one of the intermediate collimator plates in the beam line. In order to bias the conductor used for ion collection, a battery pack totaling -29~V was connected in series between the electrode and the Keithley. The stable voltage provided by the batteries allowed measurements with very little background fluctuations (down to 100~fA rms, depending on the level of noise induced by the environmental conditions). For ion counting with the MCP, one of the Moku:Go devices was programmed in oscilloscope mode to function as a pulse counter. 

The elements of the experimental setup were remote controlled using the Experimental Physics and Industrial Control System (EPICS) \cite{Dalesio94}, with the graphical user interface developed using the Control System Studio software \cite{CSSPhoebus} and some functionalities provided using Python via the \textit{PyEpics} and \textit{PCASpy} modules for implementing the Channel Access protocol. The RF amplitude of the QMF, which required precise setting for the implementation of mass scans, was stabilized using an open-loop proportional controller.    

\section{Experimental tests}
\label{tests}

\begin{figure}
    \centering
    \includegraphics[width=0.4\textwidth]{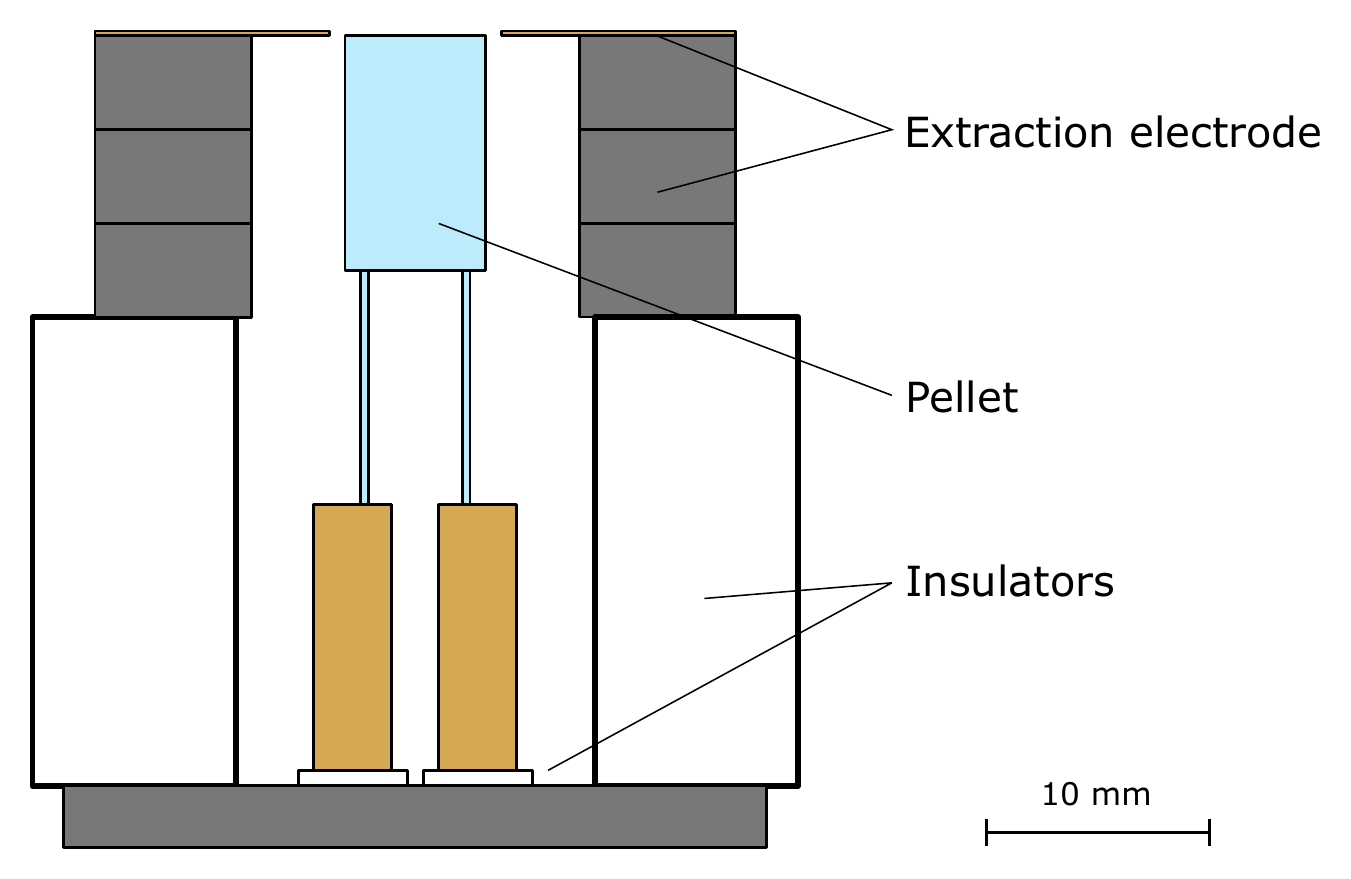}
    \caption{Schematic view of the ion source (to approximate scale) used for testing the transport of ions in the experimental setup ; a back plate is used as support for the source, on which the pellet is fixed and from which it is insulated (the insulators are represented in white); the extraction electrode surrounds the pellet to create an equipotential volume ending with a collimator plate at the same potential (top of the electrode).}
    \label{pellet}
\end{figure}

Following the assembly, cabling and alignment of the experimental setup, a series of preliminary tests were performed at IJCLab in Orsay in order to quantify its performance. An offline ion source (schematic view in Fig.~\ref{pellet}) containing a commercially available alkali pellet with $^{133}$Cs and a small quantity of $^{85,87}$Rb was used to produce the ions injected in the bRFQ. The source was mounted in front of the bRFQ, on an extension of the alignment table, at a distance close to 20~mm from the bRFQ entrance plate.  No gas cell was installed on the setup for this work and the pressure at the source position was defined by the residual gas in the bRFQ chamber. The pellet was resistively heated and concomitantly floated to 5 V, together with its acceleration electrode. The pellet allowed producing stable currents from a few 1000 per second to several nA. The ion source allowed to optimize the potentials in the beam line in order to maximize transmission to the detection chamber. The resulting optimized potentials are shown in Table~\ref{transmsetexp}. One observes that the potentials and the gradients are different from the simulated values. On the one hand, the pressure conditions were different in the experiment with respect to the simulated ones (up to a factor 5 better), as will be shown below. On the other hand, the RF matching transformers were not fine tuned and the operational frequencies were simply matched to their resonance values. With these constraints, we opted for the lowest DC voltages which could achieve satisfactory ion transmission.

\begin{table}[ht!]
\centering
\begin{tabular}{c l c}
 Section & Element & Voltage (V) \\ 
 \hline
 \multirow{4}{*}{bRFQ} 
 & RF amplitude & 141 \\
 & entrance plate & -7  \\
  & DC in & -8 \\
  & DC out & -10 \\
  \hline
   \multirow{4}{*}{mRFQ} & RF amplitude & 26.5  \\
  & DC in & -17 \\
  & DC out & -18 \\
  & exit plate & -15 \\
   \hline
  \multirow{5}{*}{QMF} & RF amplitude & 188 \\
  & entrance plate & -15 \\
  & DC in & -35 \\
  & DC out & -35 \\
  & exit plate & -30 \\
   \hline
     \multirow{4}{*}{Detection} & conductance & -125  \\
  & Lens el. 1,3 & -52 \\
  & Lens el. 2 & -281 \\
  & Grid & -28 \\
  & MCP & -1950 \\
   \hline
\end{tabular}
\caption{Optimal potentials applied to the transport line during the tests. The RF frequencies used for the bRFQ, mRFQ and QMF were of 912~kHz, 940~kHz and 732~kHz, respectively.}
\label{transmsetexp}
\end{table}

Transmission-efficiency estimates could be made by measuring the ion current on different electrodes along the beam line. Due to the presence of RF field in the close proximity of the mRFQ extraction plate or QMF entrance plate (or possible capacitive coupling between the wires in vacuum), the ion-current measurement on these elements was however impacted by electromagnetic noise. Furthermore, the -29~V bias of the battery pack used in series with the current-measurement cable was not much more negative than other neighboring potentials, which means that many of the ions could have passed through without detection, or be collected on other electrodes nearby. By consequence, no reliable ion current could be measured between the mRFQ and QMF. Nevertheless, a global transport efficiency could be determined by measuring the ion current on the bRFQ (electrodes and entrance plate connected together), the aperture plate between the QMF and the detection chamber, and the MCP grid.

The pressure conditions achieved without gas injection are given in Table \ref{pressures}.
The latter are obtained with all the pumps presented in Section~\ref{transport}, except the Edwards GXS screw pump which was not used. Instead, the nXR120i pump was used for the bRFQ chamber, while the nXR40i pump ensured the prevacuum for the turbo pumps connected to the other two chambers.
\begin{table}[!ht]
    \centering
    \begin{tabular}{c|c}
    \hline
         Chamber & Pressure (mbar) \\
    \hline
         bRFQ & $9\times10^{-3}$ \\
    \hline
         mRFQ + QMF & $\leq 5\times$ 10$^{-4}$ \\
    \hline     
         Detection & $3\times10^{-7}$ \\
    \hline
    \end{tabular}
    \caption{Pressure conditions achieved during the first commissioning of the experimental setup. The value of the pressure in the QMF chamber shows the lower measurement limit of the pressure gauge.}
    \label{pressures}
\end{table}
In these conditions, the global transmission efficiency of the ions in the whole setup was measured to be 25 $\%$. The relatively lower value compared to simulation could be due to the large source emittance (in the absence of any collimation) and the relatively low pressure in the bRFQ, which might not ensure enough radial cooling of the ions before transfer to the mRFQ. 

\begin{figure}[ht!]
\centering
\includegraphics[width=0.45\textwidth]{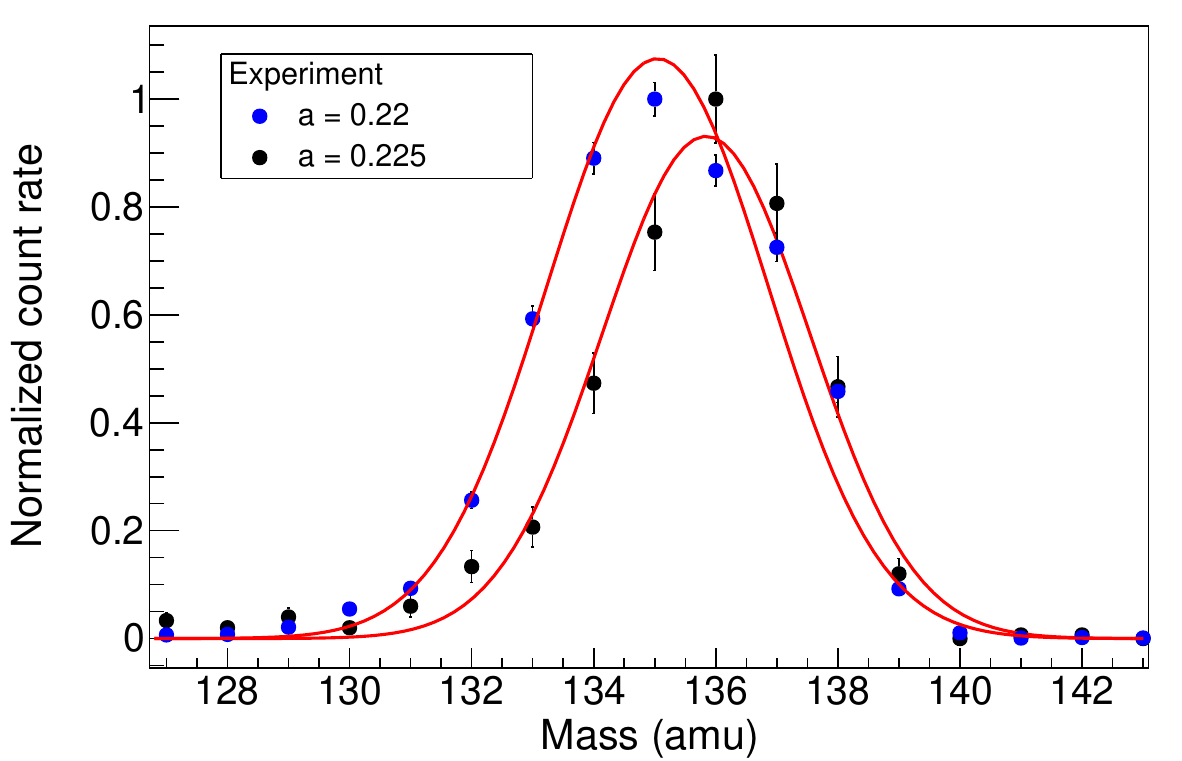}

\includegraphics[width=0.45\textwidth]{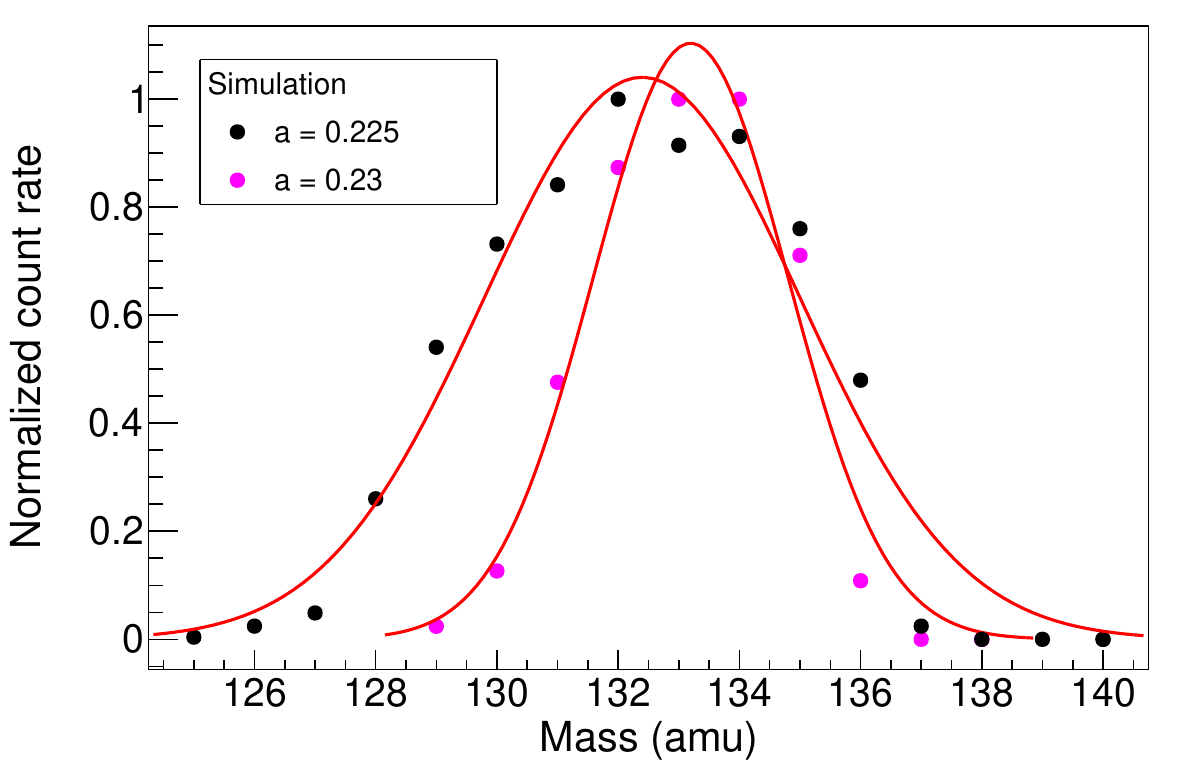}

\caption{(top) Mass scan obtained with the QMF installed in the experimental setup (RF frequency 732 kHz) compared to (bottom) a SIMION simulation for RF frequency 1~MHz). The experimental FWHM is of $4.3 \pm 0.05$~amu for $a = 0.22$ and $4.0 \pm 0.2$~amu for $a = 0.225$. The simulated FWHM is $6.1 \pm 0.3$~amu for $a=0.225$ and $3.8 \pm 0.2$~amu for $a= 0.23$.}
\label{massScans}
\end{figure}

With the optimal settings for ion transmission, the performance of the QMF was also determined and compared to SIMION simulations, as shown in Fig.~\ref{massScans}. For the simulations, the initial ion conditions and simulation space were kept the same as presented in Fig.~\ref{Simulation} and the settings used were the ones of Table~\ref{transmset}. For both measurements and simulations, the mass scan was defined in the same way as described in Section~\ref{simu}. 


The RF frequency used in the experimental tests for the QMF was 732~kHz. This leads to a measured mass resolving power of 33, on the same order as the simulated one. The shift to higher mass of the transmission peak in the actual experiment (compared to the mass of the $^{133}$Cs$^+$ ions) can be explained by the 10$\%$ asymmetry in the amplitudes of the RF phases (due to slightly different capacitive coupling of the two RF loads) which was present during the measurements. The theoretical mass represented on the x axis of the plot is calculated with the larger of the two amplitudes. This asymmetry will be corrected in the future. There is also a significant discrepancy between the ion mass and the transmission peak in the simulation due to the resolution of the SIMION mesh which leads to a slightly different effective radius of the device in the simulation space. The actual Mathieu parameter $q$ Eq.~(\ref{Mathieu}), depends quadratically on this radius and thus is very sensitive to even small deviations. 
The effective radius used in the calculation of the theoretical mass has thus been corrected for the simulation to 5.8~mm. The corrected results are presented in Fig.~\ref{massScans}, where only the displacement of the transmission maximum due to the asymmetric RF voltages can be observed in the top panel.

\begin{figure}[!ht]
\centering
\includegraphics[width = 0.48\textwidth]{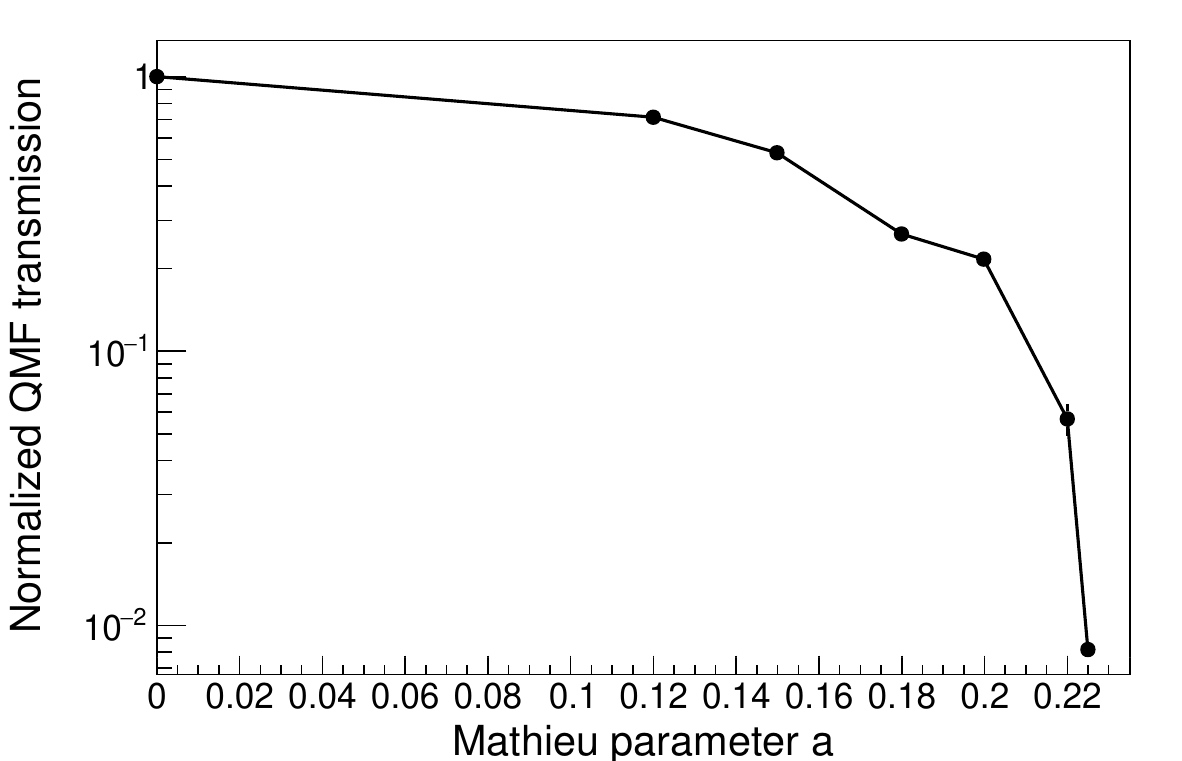}
\caption{Relative QMF transmission efficiency  as a function of the Mathieu parameter $a$}
\label{transmQMF}
\end{figure}

In general, higher resolving power of the QMF comes at the cost of lower transmission. During the tests, we have systematically varied the device resolving power (by changing the value of the Mathieu $a$ parameter) and determined the change in transmission efficiency.  Figure \ref{transmQMF} shows a measurement of the relative transmission of the QMF (with respect to the pure ion guiding mode) as a function of the Mathieu parameter $a$.

The configuration with the highest mass resolving power still leading to a usable transmission efficiency was for a Mathieu parameter $a = 0.225$. Nevertheless, a clear cut-off in transport efficiency was observed above $a = 0.22$. For the mass scan with $a = 0.22$ in Fig.~\ref{massScans}, a global transport efficiency about 1$\%$ was obtained. This relatively low efficiency is potentially due to the asymmetry between the amplitudes of the two RF phases or scattering on rest gas in the QMF. The efficiency drop is nevertheless consistent with the observed impact of the mass-filtering mode of RFQs on the transversal emittance of the emerging beams \cite{Ferrer2014b}. A careful alignment of the QMF with respect to the optics before and after, as well as a reduction of the amplitude asymmetry will allow improving this figure of merit and potentially of the resolving power. Nevertheless, the obtained efficiency is sufficient for experiments with stable ion beams, which can be produced in large quantities.  

\section{Conclusions and outlook}

In this work, we presented the FRIENDS$^3$ experimental setup, dedicated to the study of gas-cell phenomena and to the development of future gas cells for the S$^3$-LEB setup at SPIRAL2. The motivation and principles of the  FRIENDS$^3$ project were presented, as well as a detailed account of the design of the experimental setup. Simulations of ion transport were performed and showed good theoretical performance in the different operational modes, including the mass-separation mode of the QMF. 

Following assembly, first experimental tests with $^{133}$Cs$^+$ ions showed acceptable transmission of about 25$\%$ without mass separation and around $1\%$ for a resolution of 4 mass units (FWHM). We will investigate the possibility of improving the transmission of the device in high-separation mode by increasing the symmetry and stability of the applied RF field. Nevertheless, for the first tests implementing a gas cell and resonance laser ionization, the present resolving power is sufficient to separate the ions from the molecular sidebands, which will typically differ from the former by a minimum of 16 mass units (corresponding to an oxygen atom).  

Based on simulations, the setup is also capable of trapping ions in the QMF without additional gas injection, although this is yet to be verified experimentally for pressures which are reasonable within the designed pumping system. 

The experimental setup has been moved and is currently installed at GANIL, in preparation for first experiments of in-gas-cell and in-gas-jet laser ionization using a dummy gas cell. Once laser ionization and identification of photo-ions is established, different tests of ion recombination will be performed, with the aim to produce a laser-ionized beam in the gas cell, neutralize it prior to extraction and re-ionize it in the supersonic gas jet. 

The first prototype electrical gas cell is currently being manufactured and will subsequently be assembled and commissioned with the FRIENDS$^{3}$ setup, in order to study ion transport and neutralization in the presence or vicinity of an electrical field. 

\section*{CRediT authorship contribution statement}
\textbf{E.~Morin:} Methodology, Investigation, Visualization, Formal analysis, Writing - Original Draft. \textbf{W.~Dong:} Methodology, Investigation, Writing - Original Draft. \textbf{V.~Manea:} Conceptualization, Methodology, Investigation, Supervision, Funding acquisition, Writing - Original Draft. \textbf{A.~Claessens:} Methodology, Investigation. \textbf{S.~Damoy:} Resources. \textbf{R.~Ferrer:} Conceptualization, Methodology, Investigation, Resources, Writing - Review \& Editing. \textbf{S.~Franchoo:} Methodology. \textbf{S.~Geldhof:} Methodology, Resources. \textbf{T. Hourat:} Methodology, Visualization. \textbf{Yu.~Kudryavtsev:} Conceptualization, Methodology, Writing - Review \& Editing. \textbf{N.~Lecesne:} Methodology, Resources. \textbf{R. Leroy:} Resources. \textbf{D.~Lunney:} Methodology, Investigation, Resources, Writing - Review  \& Editing. \textbf{V.~Marchand:} Investigation. \textbf{E.~Minaya Ramirez:} Methodology, Resources, Writing - Review \& Editing. \textbf{S.~Raeder:} Conceptualization, Methodology. \textbf{S.~Roset:} Methodology. \textbf{Ch. Vandamme:} Software. \textbf{P.~Van den Bergh:} Methodology. \textbf{P.~Van Duppen:} Conceptualization, Methodology, Resources, Supervision.

\section*{Declaration of competing interest}

The authors declare that they have no known competing financial interests or personal relationships that could have appeared to influence the work reported in this paper.

\section*{Acknowledgments}

This work was funded by the French Research Ministry through the National Research Agency under contract number ANR-21-CE31-0001, by the French IN2P3 and GSI under the French-German collaboration agreement number PN1064, by European Union’s Horizon 2020 research and innovation programme under
Grant Agreement No. 861198-LISA-H2020-MSCA- ITN-2019, by the Research Foundation Flanders (FWO, Belgium) BOF KU Leuven (C14/22/104) and by the FWO under the Excellence of Science (EOS) program (40007501). The co-authors would like to thank the Mechanical Engineering Department of IJClab for their support in the design and manufacture of the experimental setup and the S$^3$-LEB collaboration for fruitful exchanges on the topics related to the FRIENDS$^3$ project.

\bibliographystyle{elsarticle-num} %
\bibliography{Biblio} %

\begin{thebibliography}{10}
\expandafter\ifx\csname url\endcsname\relax
  \def\url#1{\texttt{#1}}\fi
\expandafter\ifx\csname urlprefix\endcsname\relax\def\urlprefix{URL }\fi
\expandafter\ifx\csname href\endcsname\relax
  \def\href#1#2{#2} \def\path#1{#1}\fi

\bibitem{SPIRAL2}
D.~Ackermann, L.~Adoui, G.~de~Angelis, G.~Auger, T.~Aumann, F.~Azaiez, E.~Balanzat, G.~Baldacchino, M.~Barthe, E.~Bauge, P.~Bem, M.~Bender, K.~Bennaceur, J.-F. Berger, B.~Blank, J.~Blomqvist, Y.~Blumenfeld, S.~Boucard, S.~Bouffard, A.~Bracco, R.~Calabrese, B.~Cederwall, R.~Cee, P.~Chomaz, G.~Colo, M.~Colonna, D.~Curien, P.~Danielewicz, R.~Dayras, F.~de~Oliveira, M.~D. Toro, E.~Diegele, J.~Dobaczewski, T.~Ethvignot, U.~Fischer, G.~de~France, M.~Freer, U.~Garg, W.~Gelletly, A.~Gillibert, M.~Girod, S.~Goriely, H.~Goutte, J.~Grandin, F.~Gunsing, P.~Heenen, M.~Heil, K.~Heyde, S.~Hilaire, S.~Hofmann, P.~Indelicato, Z.~Janas, A.~Jokinen, J.~Jose, F.~Kaeppeler, E.~Khan, J.~Kn{\"{o}}dlseder, A.~Krasznahorkay, K.~Kratz, E.~Lamour, K.~Langanke, V.~Lapoux, E.~L. Bigot, F.~L. Blanc, X.~Ledoux, M.~Leino, S.~Lenzi, S.~Leoni, M.~Lewitowicz, D.~Lunney, E.~Maglione, A.~Maj, P.~Mantica, J.~Marques, M.~Matsuo, V.~M{\'{e}}ot, W.~Mittig, E.~Morse, O.~Naviliat-Cuncic, W.~Nazarewicz, G.~Neyens, Y.~Novikov, S.~Oberstedt, T.~Otsuka,
  F.~Parente, S.~P{\'{e}}ru, N.~Pillet, A.~Plompen, C.~Prigent, R.~Reifarth, D.~Ridikas, M.~Rivet, E.~Roeckl, M.~Rousseau, P.~Roussel-Chomaz, J.~Rozet, G.~Rudolf, K.~Rykaczewski, M.~Saint-Laurent, D.~Santonocito, J.-P. Santos, P.~Sapienza, H.~Savajols, M.~Sch{\"{a}}del, H.~Schatz, N.~Severijns, J.~Sida, F.~Sobrio, O.~Sorlin, A.~Stefanini, C.~Stodel, L.~Stuttge, I.~Testard, C.~Theisen, J.~Thomas, J.~Thomas, I.~Thompson, M.~Toulemonde, P.~van Isacker, D.~Verney, D.~Vernhet, D.~Vieira, A.~Villari, C.~Volpe, D.~Vretenar, D.~Warner, J.~Wieleczko, J.~W{\"{o}}rtche, R.~Wyss, {The Scientific Objectives of the SPIRAL2 project}, Tech. rep., GANIL (2006).
\newblock \href {https://doi.org/10.34894/VQ1DJA} {\path{doi:10.34894/VQ1DJA}}.

\bibitem{Dechery2015}
F.~D{\'{e}}chery, A.~Drouart, H.~Savajols, J.~Nolen, M.~Authier, A.~M. Amthor, D.~Boutin, O.~Delferri{\'{e}}re, B.~Gall, A.~Hue, B.~Laune, F.~{Le Blanc}, S.~Manikonda, J.~Payet, M.~H. Stodel, E.~Traykov, D.~Uriot, {Toward the drip lines and the superheavy island of stability with the Super Separator Spectrometer S3}, Eur. Phys. J. A 51~(6) (2015) 66.
\newblock \href {https://doi.org/10.1140/epja/i2015-15066-3} {\path{doi:10.1140/epja/i2015-15066-3}}.

\bibitem{Ferrer2013}
R.~Ferrer, B.~Bastin, D.~Boilley, P.~Creemers, P.~Delahaye, E.~Li{\'{e}}nard, X.~Fl{\'{e}}chard, S.~Franchoo, L.~Ghys, M.~Huyse, Y.~Kudryavtsev, N.~Lecesne, H.~Lu, F.~Lutton, E.~Mogilevskiy, D.~Pauwels, J.~Piot, D.~Radulov, L.~Rens, H.~Savajols, J.~C. Thomas, E.~Traykov, C.~V. Beveren, P.~V. den Bergh, P.~V. Duppen, {In gas laser ionization and spectroscopy experiments at the Superconducting Separator Spectrometer (S3): Conceptual studies and preliminary design}, Nucl. Instr. Meth. B 317 (2013) 570--581.
\newblock \href {https://doi.org/https://doi.org/10.1016/j.nimb.2013.07.028} {\path{doi:https://doi.org/10.1016/j.nimb.2013.07.028}}.

\bibitem{Zadvornaya2018}
A.~Zadvornaya, Characterization of the spectral resolution of the in-gas-jet laser ionization spectroscopy method, Ph.D. thesis, KU Leuven (2018).

\bibitem{Ferrer2021}
R.~Ferrer, M.~Verlinde, E.~Verstraelen, A.~Claessens, M.~Huyse, S.~Kraemer, Y.~Kudryavtsev, J.~Romans, P.~Van~den Bergh, P.~Van~Duppen, A.~Zadvornaya, O.~Chazot, G.~Grossir, V.~I. Kalikmanov, M.~Nabuurs, D.~Reynaerts, Hypersonic nozzle for laser-spectroscopy studies at 17 k characterized by resonance-ionization-spectroscopy-based flow mapping, Phys. Rev. Res. 3 (2021) 043041.
\newblock \href {https://doi.org/10.1103/PhysRevResearch.3.043041} {\path{doi:10.1103/PhysRevResearch.3.043041}}.

\bibitem{Major2005}
F.~G. Major, V.~N. Gheorghe, G.~Werth, {Charged particle traps : physics and techniques of charged particle field confinement}, Springer Berlin, Heidelberg, 2005.

\bibitem{Chauveau16}
P.~Chauveau, P.~Delahaye, G.~D. France, S.~E. Abir, J.~Lory, Y.~Merrer, M.~Rosenbusch, L.~Schweikhard, R.~N. Wolf, {PILGRIM, a Multi-Reflection Time-of-Flight Mass Spectrometer for Spiral2-S{\$}{\^{}}3{\$} at GANIL}, Nucl. Instr. Meth. B 376~(Supplement C) (2016) 211--215.
\newblock \href {https://doi.org/https://doi.org/10.1016/j.nimb.2016.01.025} {\path{doi:https://doi.org/10.1016/j.nimb.2016.01.025}}.

\bibitem{SEASON}
M.~Vandebrouck, \href{https://anr.fr/Project-ANR-20-CE31-0005}{{Spectroscopy Electron Alpha in Silicon bOx couNter}} (2020).
\newline\urlprefix\url{https://anr.fr/Project-ANR-20-CE31-0005}

\bibitem{ReyHerme2023}
E.~Rey-Herme, \href{https://theses.hal.science/tel-04206476}{{Octupole deformation in $^{221}$Ac and development of the SEASON detector}}, Phd thesis, Universit{\'{e}} Paris-Saclay (2023).
\newline\urlprefix\url{https://theses.hal.science/tel-04206476}

\bibitem{Kudryavtsev2013}
Y.~Kudryavtsev, R.~Ferrer, M.~Huyse, P.~{Van den Bergh}, P.~{Van Duppen}, The in-gas-jet laser ion source: Resonance ionization spectroscopy of radioactive atoms in supersonic gas jets, Nucl. Instr. Meth. B 297 (2013) 7--22.
\newblock \href {https://doi.org/10.1016/j.nimb.2012.12.008} {\path{doi:10.1016/j.nimb.2012.12.008}}.

\bibitem{Yang_2023}
X.~Yang, S.~Wang, S.~Wilkins, R.~G. Ruiz, Laser spectroscopy for the study of exotic nuclei, Prog. Part. Nucl. Phys. 129 (2023) 104005.
\newblock \href {https://doi.org/10.1016/j.ppnp.2022.104005} {\path{doi:10.1016/j.ppnp.2022.104005}}.

\bibitem{Ferrer2017}
R.~Ferrer, A.~Barzakh, B.~Bastin, R.~Beerwerth, M.~Block, P.~Creemers, H.~Grawe, R.~de~Groote, P.~Delahaye, X.~Fl{\'e}chard, S.~Franchoo, S.~Fritzsche, L.~Gaffney, L.~Ghys, W.~Gins, C.~Granados, R.~Heinke, L.~Hijazi, M.~Huyse, T.~Kron, Y.~Kudryavtsev, M.~Laatiaoui, N.~Lecesne, M.~Loiselet, F.~Lutton, I.~Moore, Y.~Mart{\'i}nez, E.~Mogilevskiy, P.~Naubereit, J.~Piot, S.~Raeder, S.~Rothe, H.~Savajols, S.~Sels, V.~Sonnenschein, J.-C. Thomas, E.~Traykov, C.~van Beveren, P.~van~den Bergh, P.~van Duppen, K.~Wendt, A.~Zadvornaya, {Towards high-resolution laser ionization spectroscopy of the heaviest elements in supersonic gas jet expansion}, {Nat. Commun.} 8 (2017) 14520.
\newblock \href {https://doi.org/10.1038/ncomms14520} {\path{doi:10.1038/ncomms14520}}.

\bibitem{Sonoda2013}
T.~Sonoda, M.~Wada, H.~Tomita, C.~Sakamoto, T.~Takatsuka, T.~Furukawa, H.~Iimura, Y.~Ito, T.~Kubo, Y.~Matsuo, H.~Mita, S.~Naimi, S.~Nakamura, T.~Noto, P.~Schury, T.~Shinozuka, T.~Wakui, H.~Miyatake, S.~Jeong, H.~Ishiyama, Y.~X. Watanabe, Y.~Hirayama, K.~Okada, A.~Takamine, {Development of a resonant laser ionization gas cell for high-energy, short-lived nuclei}, Nucl. Instr. Meth. B 295 (2013) 1--10.
\newblock \href {https://doi.org/10.1016/J.NIMB.2012.10.009} {\path{doi:10.1016/J.NIMB.2012.10.009}}.

\bibitem{Hirayama2017}
Y.~Hirayama, Y.~X. Watanabe, M.~Mukai, M.~Oyaizu, M.~Ahmed, H.~Ishiyama, S.~C. Jeong, Y.~Kakiguchi, S.~Kimura, J.~Y. Moon, J.~H. Park, P.~Schury, M.~Wada, H.~Miyatake, {Doughnut-shaped gas cell for KEK Isotope Separation System}, Nucl. Instr. Meth. B 412 (2017) 11--18.
\newblock \href {https://doi.org/10.1016/J.NIMB.2017.08.037} {\path{doi:10.1016/J.NIMB.2017.08.037}}.

\bibitem{Raeder2020}
S.~Raeder, M.~Block, P.~Chhetri, R.~Ferrer, S.~Kraemer, T.~Kron, M.~Laatiaoui, S.~Nothhelfer, F.~Schneider, P.~{Van Duppen}, M.~Verlinde, E.~Verstraelen, T.~Walther, A.~Zadvornaya, A gas-jet apparatus for high-resolution laser spectroscopy on the heaviest elements at ship, Nucl. Instrum. Meth. B 463 (2020) 272--276.
\newblock \href {https://doi.org/10.1016/j.nimb.2019.05.016} {\path{doi:10.1016/j.nimb.2019.05.016}}.

\bibitem{Zadvornaya2023}
A.~Zadvornaya, J.~Romero, T.~Eronen, W.~Gins, A.~Kankainen, I.~D. Moore, P.~Papadakis, I.~Pohjalainen, M.~Reponen, S.~Rinta-Antila, J.~Sar{\'{e}}n, D.~Simonovski, J.~Uusitalo, {Offline commissioning of a new gas cell for the MARA Low-Energy Branch}, Nucl. Instr. Meth. B 539 (2023) 33--42.
\newblock \href {https://doi.org/10.1016/J.NIMB.2023.03.016} {\path{doi:10.1016/J.NIMB.2023.03.016}}.

\bibitem{Ajayakumar2023}
A.~Ajayakumar, J.~Romans, M.~Authier, Y.~Balasmeh, A.~Brizard, F.~Boumard, L.~Caceres, J.~F. Cam, A.~Claessens, S.~Damoy, P.~Delahaye, P.~Desrues, W.~Dong, A.~Drouart, P.~Duchesne, R.~Ferrer, X.~Fl{\'{e}}chard, S.~Franchoo, P.~Gangnant, S.~Geldhof, R.~P. de~Groote, F.~Ivandikov, N.~Lecesne, R.~Leroy, J.~Lory, F.~Lutton, V.~Manea, Y.~Merrer, I.~Moore, A.~Ortiz-Cortes, B.~Osmond, J.~Piot, O.~Pochon, S.~Raeder, A.~de~Roubin, H.~Savajols, D.~Studer, E.~Traykov, J.~Uusitalo, C.~Vandamme, P.~{Van den Bergh}, P.~{Van Duppen}, K.~Wendt, {In-gas-jet laser spectroscopy with S3-LEB}, Nucl. Instr. Meth. B 539 (2023) 102--107.
\newblock \href {https://doi.org/10.1016/J.NIMB.2023.03.020} {\path{doi:10.1016/J.NIMB.2023.03.020}}.

\bibitem{Lantis2024}
J.~Lantis, A.~Claessens, D.~M\"unzberg, J.~Auler, M.~Block, P.~Chhetri, C.~E. D\"ullmann, R.~Ferrer, F.~Giacoppo, M.~J. Guti\'errez, F.~Ivandikov, O.~Kaleja, T.~Kieck, E.~Kim, M.~Laatiaoui, N.~Lecesne, V.~Manea, S.~Nothhelfer, S.~Raeder, J.~Romans, E.~Romero-Romero, A.~de~Roubin, H.~Savajols, S.~Sels, M.~Stemmler, P.~Van~Duppen, T.~Walther, J.~Warbinek, K.~Wendt, A.~Yakushev, A.~Zadvornaya, In-gas-jet laser spectroscopy of $^{254}\mathrm{No}$ with jetris, Phys. Rev. Res. 6 (2024) 023318.
\newblock \href {https://doi.org/10.1103/PhysRevResearch.6.023318} {\path{doi:10.1103/PhysRevResearch.6.023318}}.

\bibitem{Kudryavtsev2016}
Y.~Kudryavtsev, et~al., {A new in-gas-laser ionization and spectroscopy laboratory for off-line studies at KU Leuven}, Nucl. Instrum. Meth. B 376 (2016) 345--352.
\newblock \href {https://doi.org/10.1016/J.NIMB.2016.02.040} {\path{doi:10.1016/J.NIMB.2016.02.040}}.

\bibitem{FRIENDS3}
V.~Manea, \href{https://anr.fr/Project-ANR-21-CE31-0001}{{Fast Radioactive Ion Extraction and Neutralization Device for S3 (FRIENDS3)}} (2021).
\newline\urlprefix\url{https://anr.fr/Project-ANR-21-CE31-0001}

\bibitem{Dong2024}
W.~Dong, {Developments for the laser spectroscopy of exotic nuclei with the S$^3$ Low Energy Branch and the FRIENDS$^3$ project}, Phd thesis, Universit{\'{e}} Paris-Saclay (2024).
\newblock \href {https://doi.org/10.34894/VQ1DJA} {\path{doi:10.34894/VQ1DJA}}.

\bibitem{Lautenschlager2016}
F.~Lautenschl{\"{a}}ger, P.~Chhetri, D.~Ackermann, H.~Backe, M.~Block, B.~Cheal, A.~Clark, C.~Droese, R.~Ferrer, F.~Giacoppo, S.~G{\"{o}}tz, F.~P. He{\ss}berger, O.~Kaleja, J.~Khuyagbaatar, P.~Kunz, A.~K. Mistry, M.~Laatiaoui, W.~Lauth, S.~Raeder, T.~Walther, C.~Wraith, {Developments for resonance ionization laser spectroscopy of the heaviest elements at SHIP}, Nucl. Instr. Meth. B 383 (2016) 115--122.
\newblock \href {https://doi.org/10.1016/J.NIMB.2016.06.001} {\path{doi:10.1016/J.NIMB.2016.06.001}}.

\bibitem{kudryavtsev2001gas}
Y.~Kudryavtsev, et~al., {A gas cell for thermalizing, storing and transporting radioactive ions and atoms. Part I: Off-line studies with a laser ion source}, Nucl. Instr. Meth. B 179~(3) (2001) 412--435.

\bibitem{Facina2004}
M.~Facina, B.~Bruyneel, S.~Dean, J.~Gentens, M.~Huyse, Y.~Kudryavtsev, P.~{Van Den Bergh}, P.~{Van Duppen}, {A gas cell for thermalizing, storing and transporting radioactive ions and atoms. Part II: On-line studies with a laser ion source}, Nucl. Instr. Meth. B 226~(3) (2004) 401--418.
\newblock \href {https://doi.org/10.1016/J.NIMB.2004.06.031} {\path{doi:10.1016/J.NIMB.2004.06.031}}.

\bibitem{Moore2010}
I.~D. Moore, T.~Kessler, T.~Sonoda, Y.~Kudryavstev, K.~Per{\"{a}}j{\"{a}}rvi, A.~Popov, K.~D. Wendt, J.~{\"{A}}yst{\"{o}}, {A study of on-line gas cell processes at IGISOL}, Nucl. Instr. Meth. B 268~(6) (2010) 657--670.
\newblock \href {https://doi.org/10.1016/J.NIMB.2009.12.001} {\path{doi:10.1016/J.NIMB.2009.12.001}}.

\bibitem{NUBASE2020}
F.~G. Kondev, M.~Wang, W.~J. Huang, S.~Naimi, G.~Audi, {The NUBASE2020 evaluation of nuclear physics properties *}, Chinese Phys. C 45~(3) (2021) 30001.
\newblock \href {https://doi.org/10.1088/1674-1137/abddae} {\path{doi:10.1088/1674-1137/abddae}}.

\bibitem{Laatiaoui16}
M.~Laatiaoui, W.~Lauth, H.~Backe, M.~Block, D.~Ackermann, B.~Cheal, P.~Chhetri, C.~E. D{\"{u}}llmann, P.~{Van Duppen}, J.~Even, R.~Ferrer, F.~Giacoppo, S.~G{\"{o}}tz, P.~He{\ss}berger, M.~Huyse, O.~Kaleja, J.~Khuyagbaatar, P.~Kunz, F.~Lautenschl{\"{a}}ger, A.~K. Mistry, S.~Raeder, E.~{Minaya Ramirez}, T.~Walther, C.~Wraith, A.~Yakushev, {Atom-at-a-time laser resonance ionization spectroscopy of nobelium}, Nature 538~(7626) (2016) 495--498.
\newblock \href {https://doi.org/10.1038/nature19345} {\path{doi:10.1038/nature19345}}.

\bibitem{Dong2025}
W.~Dong, et~al., A prototype gas cell for the stopping, extraction and neutralization of radioactive nuclei at spiral2-s$^3$, in preparation (2025).

\bibitem{Claessens2024}
A.~Claessens, {Laser ionization spectroscopy of $^{254}$No and $^{229}$Th in hypersonic gas jets}, Phd thesis, KU Leuven (2024).

\bibitem{Dahl2000}
D.~A. Dahl, simion for the personal computer in reflection, Int. Journ. Mass Spectrom. 200~(1) (2000) 3--25, volume 200: The state of the field as we move into a new millenium.
\newblock \href {https://doi.org/https://doi.org/10.1016/S1387-3806(00)00305-5} {\path{doi:https://doi.org/10.1016/S1387-3806(00)00305-5}}.

\bibitem{Manura2007}
D.~Manura, Additional notes on the simion hs1 collision model, Tech. rep., Scientific Intrument Services, Inc. (2007).

\bibitem{Dalesio94}
L.~R. Dalesio, J.~O. Hill, M.~Kraimer, S.~Lewis, D.~Murray, S.~Hunt, W.~Watson, M.~Clausen, J.~Dalesio, {The experimental physics and industrial control system architecture: past, present, and future}, Nucl. Instr. Meth. A 352~(1-2) (1994) 179--184.
\newblock \href {https://doi.org/10.1016/0168-9002(94)91493-1} {\path{doi:10.1016/0168-9002(94)91493-1}}.

\bibitem{CSSPhoebus}
\href{https://controlssoftware.sns.ornl.gov/css{\_}phoebus/}{{CS-Studio (Phoebus)}}.
\newline\urlprefix\url{https://controlssoftware.sns.ornl.gov/css{\_}phoebus/}

\bibitem{Ferrer2014b}
R.~Ferrer, A.~Kwiatkowski, G.~Bollen, D.~Lincoln, D.~Morrissey, G.~Pang, R.~Ringle, J.~Savory, S.~Schwarz, Ion beam properties after mass filtering with a linear radiofrequency quadrupole, Nucl. Instrum. Meth A 735 (2014) 382--389.
\newblock \href {https://doi.org/https://doi.org/10.1016/j.nima.2013.09.054} {\path{doi:https://doi.org/10.1016/j.nima.2013.09.054}}.

\end{thebibliography}

\end{document}